\documentclass[sigconf]{acmart}
\usepackage[utf8x]{inputenc}
\usepackage{tabularx}
\restylefloat{figure} 
\AtBeginDocument{%
  \providecommand\BibTeX{{%
    \normalfont B\kern-0.5em{\scshape i\kern-0.25em b}\kern-0.8em\TeX}}}

\copyrightyear{2020}
\acmYear{2020}
\setcopyright{rightsretained}
\acmConference[CUI '20]{2nd Conference on Conversational User Interfaces}{July 22--24, 2020}{Bilbao, Spain}
\acmDOI{10.1145/3405755.3406136}
\acmISBN{978-1-4503-7544-3/20/07}

\begin{document}

%%
%% The "title" command has an optional parameter,
%% allowing the author to define a "short title" to be used in page headers.
\title{Transparency in Language Generation: Levels of Automation}

%%
%% The "author" command and its associated commands are used to define
%% the authors and their affiliations.
%% Of note is the shared affiliation of the first two authors, and the
%% "authornote" and "authornotemark" commands
%% used to denote shared contribution to the research.
\author{Justin Edwards}
\affiliation{justin.edwards@ucdconnect.ie}

\author{Allison Perrone}
\affiliation{allison@batcamp.org}

\author{Philip R. Doyle}
\affiliation{philip.doyle1@ucdconnect.ie}

%%
%% By default, the full list of authors will be used in the page
%% headers. Often, this list is too long, and will overlap
%% other information printed in the page headers. This command allows
%% the author to define a more concise list
%% of authors' names for this purpose.
% \renewcommand{\shortauthors}{Bleakley, Wu, Pandey and Edwards}

%%
%% The abstract is a short summary of the work to be presented in the
%% article.
\begin{abstract}
 Language models and conversational systems are growing increasingly advanced, creating outputs that may be mistaken for humans. Consumers may thus be misled by advertising, media reports, or vagueness regarding the role of automation in the production of language. We propose a taxonomy of language automation, based on the SAE levels of driving automation, to establish a shared set of terms for describing automated language. It is our hope that the proposed taxonomy can increase transparency in this rapidly advancing field.

\end{abstract}

%%
%% The code below is generated by the tool at http://dl.acm.org/ccs.cfm.
%% Please copy and paste the code instead of the example below.
%%



%%
%% Keywords. The author(s) should pick words that accurately describe
%% the work being presented. Separate the keywords with commas.

\maketitle

\section{Introduction}

Increasingly, transparency is becoming a key practical and ethical concern for developers of natural language technologies {\cite{candello_cuichi_2020}}. Conversational research suggests knowledge about the type of partner a person is talking to (human or computer) has a significant impact: on language choices in dialogue \cite{cowan_whats_2019,cowan_does_2015}; on perceptions of a partner’s knowledge and capabilities \cite{doyle_mapping_2019}; on perceptions of interpersonal connection \cite{doyle_mapping_2019, purington_alexa_2017}; and on perceptions of trustworthiness \cite{torre_trust_2018}. Failure to disclose when someone is talking to a computer rather than a person can also lead to heightened expectations about system capability. Subsequent failure to meet these expectations can lead to frustration, limited use and even abandonment of CUIs \cite{luger_like_2016}. Transparency is also a key element of recent policy implementations such as GDPR, which guarantees European citizens a right to transparency in their interactions with technology \cite{european_2016}. Here, we propose a structure for 'levels of automation' that can be used to clearly delineate the roles of humans and machines in generating output. Our hope is that the levels of automation posited here - inspired by the SAE taxonomy of driving automation - will be a first step toward greater transparency in the field of conversational technology, inspiring iterative refinement of this taxonomy that produces universal descriptions for these technologies. 

\section{SAE Levels}

Issued in 2014, the SAE taxonomy of driving automation levels clearly categorizes automated driving systems according to the level of control possessed by the driver and the vehicle, respectively. The clarity of language afforded by the SAE taxonomy allows for marketing of new technologies that avoids misleading consumers \cite{barry_levels_2019} as well as allowing researchers, policymakers and journalists to discuss emerging technologies \cite{milakis_long-term_2019}. It also provides consumers with clarity around levels of automation and control. Table 1 presents a summarized version of the SAE levels of driving automation.

\vspace*{-\baselineskip}
\newcolumntype{s}{>{\hsize=.22\hsize}X}
\begin{table}[h]
\caption{SAE Levels of driving automation}
\label{Table 1:}
\begin{tabularx}{\linewidth}{s|X}
SAE Level & Description                                                                                                                                   \\ \hline
0         & No automation                                                                                                                                 \\ \hline
1         & Driver assistance (a single automated system like cruise control)                                                                             \\ \hline
2         & Partial automation (vehicle can operate autonomously, but human monitors the environment and can take control at any time)                    \\ \hline
3         & Conditional automation (vehicle can monitor environment, operate autonomously, but human must be available to takeover in some situations) \\ \hline
4         & High automation (under certain circumstances, the vehicle is fully autonomous, human takeover is option in other circumstance)      \\ \hline
5         & Full automation: (No human interaction required, takeover may be disabled)                                                                   
\end{tabularx}
\end{table}

SAE levels are defined in terms of the role and responsibility of the driver in relation to the vehicle's automated features. As such, they give insight into varying levels of human/system control across different in-car automation systems and across various driving events a system may or may not be designed to handle. As seen in Table 1, human operators are in control of levels 2 and below, whilst the automated driving system is in control at levels 3 and above. This delineation helps in setting appropriate expectations for drivers, aids policymakers in establishing appropriate legal frameworks, and allows for greater accuracy in reporting when these human-computer interactions fail. The SAE levels were developed through committee discussion among driving engineers and industry stakeholders to develop a common set of terms for this specific issue. It is our hope that by structuring similar levels in the field of language generation, we too can give a common language to our research community, enhancing transparency in the field.

\section{Levels of Language Generation Automation}

Like the SAE levels, we propose a structure of 6 levels of automation that define the role of a human author when supported by automated language generation systems, including those using both rule-based and probabilistic approaches. Below, we posit definitions and examples of each of the six levels of language generation.

\textbf{Level 0: Fully human-written language: } indicative of language written and selected exclusively by a human. While editing assistance like spell check may be used at this level, lexical choices are entirely controlled by a human.

\textbf{Level 1: Language assistance:} includes language that is entirely written and selected by a human, but may be scripted to present automatically. These may include highly constrained chatbots designed for a specific role, or phone trees limited by predetermined sequences. Systems like this present users with a limited set of dialogue options per turn, rather than allowing a user to freely enter language. Dialogues may take varied branching paths, but only within the confines of predetermined sets of commands and responses that were generated by a human author. No novel language is generated throughout these interactions. In this way, level 1 automation allows for automation of scripted interactions rather than automation of novel generation.

\textbf{Level 2: Partial automation:} includes language generated\\ through shared effort between a human author and an automated system. This may include language written by a human then selected algorithmically, and/or language generated algorithmically then selected by a human. Many modern Twitterbots take the former approach, including Twitterbots built using Tracery (e.g. Lost Tesla Bot\footnote{twitter.com/LostTesla}. Tracery is a language generation approach that generates text through slot filling using random or conditional selection \cite{compton_tracery_2015}. An author can create a number of templates, called origins, and define slots and lists of potential entries for each slot. A Twitterbot employing Tracery thus produces novel text by combining several human-written texts.

An example of partial automation can also be seen in output from Voicebox, a predictive text tool by Botnik Studios\footnote{botnik.org}. Voicebox is a Markov chain-based predictive text keyboard that can be trained on existing text uploaded by the author (e.g.. a compilation of an artist’s lyrics, a body of text from a novel). The author then generates a new text by selecting one word at a time. Level 2 partial automation still requires a high level of human involvement as language must be initially provided by a human author, but novel generation is possible.

\textbf{Level 3: Conditional automation:} Here language generation is accomplished by automation, with human effort reserved primarily for selecting the generated language for use. At this level, the role of authorship is shared between a human and a machine. As an example, in a recent article in The Economist, contributor Tom Standage interviewed GPT-2 by inputting interview questions about technology to watch in 2020 \cite{standage_artificial_2019}. GPT-2, a large stochastic language model \cite{radford_language_2019}, was instantiated in a shared writing environment in which an author could input text cues for GPT-2 to continue to generate language. Standage authored questions, received several generated answers, and selected which answers to publish. Although a higher level of automation was asserted, this masks the role Standage had in composing the interview, particularly in selecting which answers to use. An interaction like this is more accurately described as a Level 3, conditional automation involving a shared task between a human and a machine.

\textbf{Level 4: High automation: } Level 4 language automation requires no human supervision. Here, language is generated and selected stochastically, though constrained to a specific domain or language task. One currently available implementation of level 4 language automation is AI Dungeon, a GPT-2 based text adventure role-playing game\footnote{aidungeon.io}. Gameplay in AI Dungeon is generated by an instantiation of GPT-2 trained on text from role-playing games, with the game responding to human text input with stochastically generated story development, creating emergent gameplay between the game and player. Similar implementations could be used in a variety of other use-cases, like educational tools to encourage children to write, and marketing tools such as embodied agents promoting brand engagement. While training models and defining domains may require high human effort, language generation is not performed by a human at this level of automation. This fully automatised generation differentiates level 4 from level 3, while domain constraints differentiate it from level 5.

\textbf{Level 5: Fully automated language: }Level 5 represents fully automated language generation. Like level 4, fully automated language requires no human supervision and is both produced and selected stochastically. However, unlike level 4, language is not constrained to a particular domain, task, or topic. This is a long-held goal for general artificial intelligence \cite{nilsson_quest_2010} and has not yet been accomplished. To merit consideration as a level 5 automated language system, a system would need to be capable, without modification, of various language tasks including both task-oriented transactional dialogue and open-ended social dialogue. 

Level 5 systems may be most useful as a template that users could then constrain for different specific purposes, thus rendering specific implementations into level 4 systems. While there is some evidence that people enjoy chatting with natural-sounding chatbots \cite{shum_eliza_2018}, other work casts doubt suggesting these preferences may be due to the novelty of the interaction \cite{clark_what_2019}, which may mean there are limited use-cases for this level of automation. It should be made clear that the level of automation of a language system does not necessarily correspond to the quality of outputs nor to the utility of the system overall.

\section{Conclusion}
Conversational systems and language generation tools are becoming increasingly advanced, blurring lines between human-generated and computer-generated language. Degrees of automation have been clearly delineated in the field of automated driving though the use of a shared set of definitions that can be understood by a variety of stakeholders. By using a similar taxonomy in the field of conversational technology, we can ensure that this field maintains transparency in discussion of generated language, ensuring consumers are not misled when interacting with these systems.

\bibliographystyle{ACM-Reference-Format}
\bibliography{robots}

\end{document}